\documentclass[reprint,twocolumns,aps,prl]{revtex4-1}

\usepackage{color,rotating,graphicx}
\usepackage{amsmath,ulem}
\usepackage{dsfont}

\newcommand{\ri}{{\rm i}}
\newcommand{\re}{{\rm e}}
\newcommand{\rd}{{\rm d}}

\newcommand{\kb}{k_{\rm B}}

\begin{document}

\title{Breakdown of detailed balance for thermal radiation by synthetic fields}

\author{S.-A. Biehs}
\email{s.age.biehs@uni-oldenburg.de}
\affiliation{Institut f\"{u}r Physik, Carl von Ossietzky Universit\"{a}t, D-26111 Oldenburg, Germany}
\author{G. S. Agarwal}
\email{girish.agarwal@tamu.edu}
\affiliation{ Institute for Quantum Science and Engineering and Department of Biological and 
Agricultural Engineering Department of Physics and Astronomy, Texas A \& M University, College Station, Texas 77845, USA}
%
%

%\date{\today}

%\pacs{44.40.+a, 78.20.N-, 03.50.De, 66.70.-f}
\begin{abstract}
In recent times the possibility of non-reciprocity in heat transfer between two bodies has been extensively studied. In particular the role of strong magnetic fields has been investigated. A much simpler approach with considerable flexibility would be to consider heat transfer in synthetic electric and magnetic fields which are easily applied. We demonstrate the breakdown of detailed balance for the heat transfer function $\mathcal{T}(\omega)$, i.e.\ the spectrum of heat transfer between two objects due to the presence of synthetic electric and magnetic fields. The spectral measurements carry lot more physical information and were the reason for the quantum theory of radiation. We demonstrate explicitly the synthetic field induced non-reciprocity in the heat transfer transmission function between two graphene flakes and for the Casimir coupling between two objects. Unlike many other cases of heat transfer, the latter case has interesting features of the strong coupling. Further the presence of synthetic fields affects the mean occupation numbers of two membranes and propose this system for the experimental verification of the breakdown of detailed balance.
\end{abstract}

\maketitle
%%%%%%%%%%%%%%%%%%%%%%%%%%%%%%%%%%%%%%%%%%%%%%%
%
% Introduction
%
%%%%%%%%%%%%%%%%%%%%%%%%%%%%%%%%%%%%%%%%%%%%%%%
%
% heat flux modulation (modulation, shuttling, etc.)
% non-reciprocal heat flux (broken time reversal symmetry)
% synthetic fields -> non-reciprocal energy flux
% show: non-reciprocal spectra but reciprocal energy flux
% Two systems: NFRHT between two graphene sheets; Casimir force coupled membranes

Reciprocity and detailed balance are at the heart of Kirchhoff's law stating that the absorptivity equals emissivity for any frequency and
angle of incidence. In fact, the second law of thermodynamics enforces the reciprocity or better detailed balance of the radiative heat
transfer between two objects. Here it is unimportant if far-field heat transfer where Planck's blackbody determines the upper limit is considered or 
near-field heat transfer where the blackbody limit is not a limit anymore~\cite{ReviewCarlos,Limit,RMP} as experimentally tested by a great number of
experiments~\cite{HuEtAl2008,OttensEtAl2011, Kralik2012, LimEtAl2015, WatjenEtAl2016,BernadiEtAl2016,SongEtAl2016, FiorinoEtAl2018a, FiorinoEtAl2018b} within the last decade. How the second law enforces detailed balance can be understood~\cite{Latella2017} by considering the heat 
flux between two objects by first taking the transferred power from object $a$ to object $b$
\begin{equation}
	P_{a \rightarrow b} = \int_0^\infty \frac{\rd \omega}{2 \pi} \hbar \omega n_a(\omega) \mathcal{T}_{ab}(\omega)
\end{equation}
where $\hbar$ is the Planck constant, $n_a(\omega) = (\exp(\hbar\omega/\kb T_a) - 1)^{-1}$ is the photonic occupation number, $\kb$ is the Boltzmann constant, and $T_a$ is the temperature of object $a$. The quantity  $\mathcal{T}_{ab}(\omega)$ is a heat transfer function (HTF) for the heat flow from object $a$ to object $b$. Similarly, the heat flow from object $b$ to object $a$ is given by
\begin{equation}
	P_{b \rightarrow a} = \int_0^\infty \frac{\rd \omega}{2 \pi} \hbar \omega n_b(\omega) \mathcal{T}_{ba}(\omega)
\end{equation}
with $n_b(\omega) = (\exp(\hbar\omega/\kb T_b) - 1)^{-1}$ and $T_b$ the temperatur of object $b$. In thermal equilibrium the objects have the same temperature $T_a = T_b$ and therefore there is no net heat flow which means that $P_{a \rightarrow b} = P_{b \rightarrow a}$ and hence
\begin{equation}
	\int_0^\infty \frac{\rd \omega}{2 \pi} \hbar \omega n_a(\omega) \bigl[\mathcal{T}_{ba}(\omega) - \mathcal{T}_{ab}(\omega)\bigr] = 0.
	\label{Eq:Integral}
\end{equation}
Since this expression holds for any value of temperature $T_a = T_b$ and therefore for different spectral weighting by $n_a$ it can be 
concluded that the validity of the second law of thermodynamics is equivalent to the relation $\mathcal{T}_{ab}(\omega) = \mathcal{T}_{ba}(\omega)$ regardless of any symmetry~\cite{FanRec}. That means that even when time reversal symmetry is broken by applying a magnetic field or using topological Weyl semi-metals, for instance, detailed balance of the energy HTF must be fulfilled. However, in non-reciprocal systems the detailed balance of thermal radiation can be nearly completely violated when considering three objects~\cite{ZhuEtAl2014} which also offers applications for optimized non-reciprocal thermo-photovoltaic energy conversion~\cite{ParkEtAl2022}. 
Similarly, the HTFs do not need to fulfill $\mathcal{T}_{ab}(\omega) = \mathcal{T}_{ba}(\omega)$ when at least a third object $c$ or a non-reciprocal environment are present. In such many-body systems therefore several interesting effects for thermal radiation in general and radiative heat exchange in nanoparticle systems~\cite{RMP} in particular could be highlighted like persistent heat currents and 
heat fluxes~\cite{zhufan,Silveirinha,CircularHeatFlux}, persistent spin and angular momenta~\cite{Silveirinha,CircularHeatFlux,Zubin2019},
giant thermal resistance~\cite{Latella2017,He2020}, a normal and anomalous Hall effect for thermal radiation~\cite{hall,OttEtAl2019,ahall}, as 
well as a diode effect with non-reciprocal surface waves~\cite{NRdiode}. In all these studied systems, in order to realize a non-reciprocal heat flux or a violation of detailed balance the presence of a third body seems to be a necessary condition. However, within the framework of fluctuational electrodynamics and the scattering formalism~\cite{Krueger1,Krueger2} a formal proof detailed in Ref.~\cite{Herz} shows that $\mathcal{T}_{ab}(\omega) = \mathcal{T}_{ba}(\omega)$ if the environment and the objects fulfill both Lorentz reciprocity~\cite{Alu}. Therefore in principle for radiative heat transfer between two objects with non-reciprocal properties in a reciprocal environment detailed balance can be broken even though in practive this has not been observed so far.

Interestingly, the presence of synthetic electric and magnetic fields offers the possibility to break the detailed balance of energy transmission functions even for only two coupled resonators which results in a non-reciprocal energy transmission as shown theoretically and verified experimentally~\cite{PetersonEtAl2019}. The synthetic fields are generated by external modulation of the resonance frequency of the two resonators which first of all generates side bands which can be understood by the presence of a synthetic electric field in the synthetic frequency domain~\cite{YuanEtAl2018}. When the modulation of the two resonators is phase-shifted a synthetic magnetic field for the photons is generated~\cite{TzuanEtAl2014} which enables the Aharonov-Bohm effect for photons~\cite{FangEtAl2012}, for instance. Now, dynamic modulations of temperatures or material properties have also been considered for modulation of radiative heat exchange between two or more objects~\cite{ZwolVo2,Ito,Minnich,Minnich2} showing that the modulation of the temperature or chemical potential can result in a shuttling effect~\cite{Latella} and the modulation of material properties can be used to modulate the radiative heat flux between two or more objects~\cite{Cuevas,Messina}. However, in all those works the HTF for the radiative heat exchange between two bodies are again strictly fulfilling detailed balance, i.e.\  $\mathcal{T}_{ab}(\omega) = \mathcal{T}_{ba}(\omega)$. 

In this letter, by using a quantum Langevin equation approach to treat heat transfer we show that synthetic fields can lead to a breakdown of detailed balance for the HTF between two resonant objects, i.e.\ we explicitely show that $\mathcal{T}_{ab}(\omega) \neq \mathcal{T}_{ba}(\omega)$. We further show that this broken detailed balance does not result in a non-reciprocal heat flux, i.e.\ we still have $P_{a \rightarrow b} = P_{b \rightarrow a}$ and the validity of Eq.~(\ref{Eq:Integral}). We will discuss these features for the radiative heat flux between two graphene flakes in which case the synthetic fields are realized by modulating Fermi energies. Furthermore, we propose to measure the broken detailed balance in the strong-coupling regime of two Casimir-force coupled membranes as used in recent experiments like in Ref.~\cite{FongEtAl}.  

%%%%%%%%%%%%%%%%%%%%%%%%%%%%%%%%%%%%%%%%%%%%%%
%
% Model
%
%%%%%%%%%%%%%%%%%%%%%%%%%%%%%%%%%%%%%%%%%%%%%%

In the following we describe the near-field radiative heat flux between two graphene flakes as well as Casimir force coupled membranes by two coupled oscillators~\cite{JOSA,Barton,Karthik}. The oscillator frequencies $\omega_{a/b}$ then correspond to the frequencies of the main optical or vibrational modes of the graphene flakes or the membranes and their damping is described by the damping constants $\kappa_{a/b}$. The coupling strength between the oscillators $g$ quantifies the interaction strength of the graphene flakes or membranes due to the fluctuational electromagnetic fields which are at the origin of the radiative heat transfer and Casimir force. Then the coupled oscillators can be described by a set of two quantum Langevin equations~\cite{AgarwalBook2012,Suppl}
\begin{align}
	\dot{a} &= - \ri \omega_a a - \kappa_a a - \ri g b + F_a, \\
	\dot{b} &= - \ri \omega_b b - \kappa_b b - \ri g a + F_b 
\end{align}
for the lowering operators $a$ and $b$ of the two coupled oscillators.  Furthermore, both oscillators are assumed to be coupled to their own baths which enter here through the bath operators $F_{a/b}$ into the description. 
%The bath operators fulfill the correlation functions
%\begin{equation}
%	\langle F_{a/b}^\dagger F_{a/b} \rangle  = 2 \kappa_{1/2} n_{a/b}
%\end{equation}
%where the mean occupation numbers of the oscillators  
%\begin{equation}
%	n_{a/b} = \frac{1}{\re^{\hbar \omega_{a/b}/ \kb T_{a/b}} - 1}
%\end{equation}
%are determined by the oscillator frequencies $\omega_{a/b}$ and the bath temperatures $T_{a/b}$. 

\begin{figure}
	\centering
	  \includegraphics[width=0.45\textwidth]{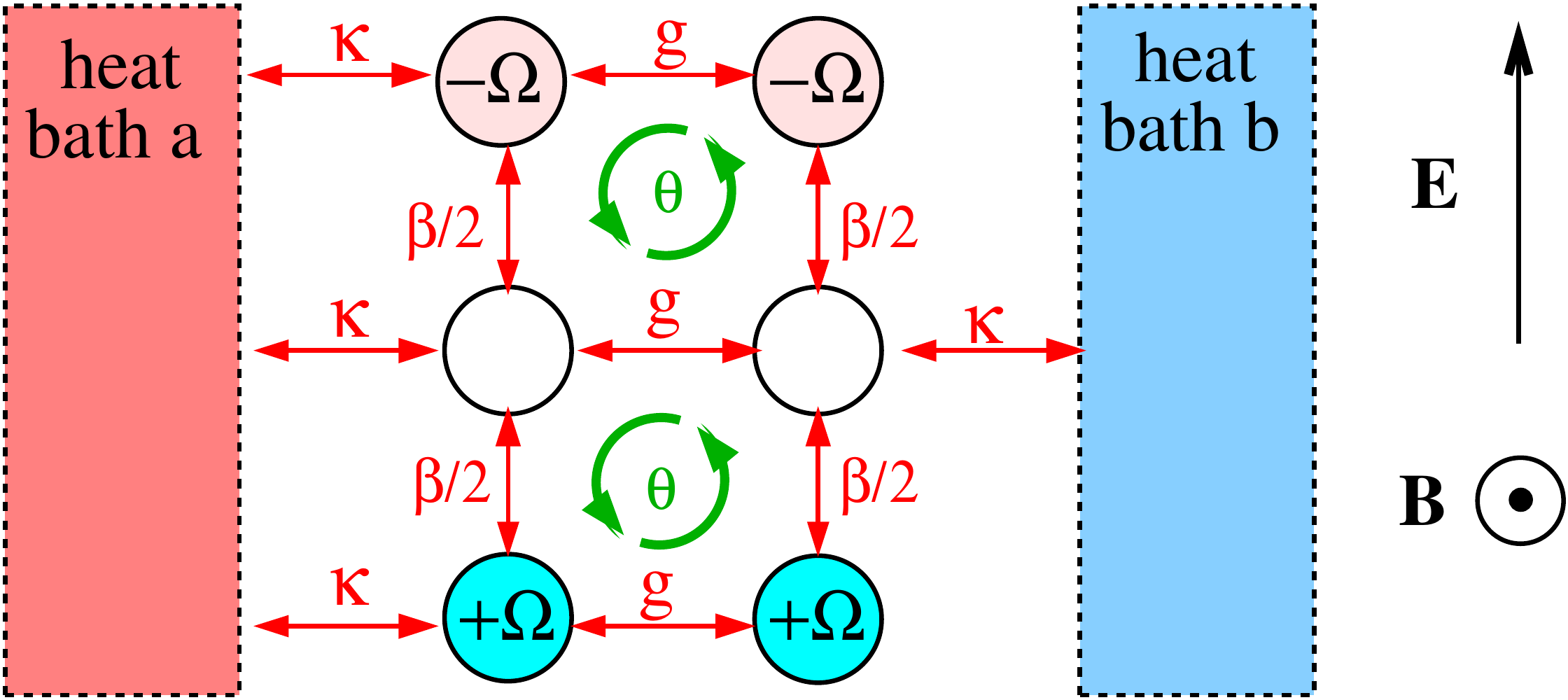}
	\caption{\label{Fig:Sketch} Sketch of the forward heat flux $P_{a \rightarrow b}$ in the considered two couples oscillators with periodic modulation in the synthetic dimension with the synthetic electric and magnetic fields $\mathbf{E}$ and $\mathbf{B}$.}
\end{figure}

Now, we introduce synthetic electric and magnetic fields via the identical frequency modulation of both oscillators 
\begin{equation}
	\omega_{a} \rightarrow \omega_a + \beta \cos(\Omega t) \quad\text{and}\quad  \omega_{b} \rightarrow \omega_b + \beta \cos(\Omega t + \theta) \label{Eq:omegaa} 
\end{equation}
with modulation frequency $\Omega$, amplitude $\beta$ and with a phase shift $\theta$. By Fourier transforming the coupled Langevin equations into frequency space we obtain the set of equations in the compact form
\begin{equation}
	\boldsymbol{\psi} = \mathds{M} \mathbf{F} + \frac{\beta}{2 \ri} \mathds{M} \mathds{Q}_+ \boldsymbol{\psi}_+ + \frac{\beta}{2 \ri} \mathds{M} \mathds{Q}_- \boldsymbol{\psi}_- 
\label{Eq:coupledLangevin}
\end{equation}
by introducing the vectors $\boldsymbol{\psi} = \bigl( a(\omega) , b(\omega) \bigr)^t$, $\boldsymbol{\psi}_\pm = \bigl( a(\omega \pm \Omega) , b(\omega \pm \Omega) \bigr)^t$, and $\mathbf{F} = \bigl(F_a(\omega), F_b(\omega) \bigr)^t$, 
and the matrices
\begin{equation}
	\mathds{M} = \mathds{A}^{-1} \quad \text{with} \quad \mathds{A} = \begin{pmatrix} X_a & \ri g \\ \ri g & X_b \end{pmatrix}
\end{equation}
so that
\begin{equation}
	\mathds{M} =  \frac{1}{X_a X_b + g^2}\begin{pmatrix} X_b & - \ri g \\ - \ri g & X_a \end{pmatrix}
\end{equation}
introducing $X_{a/b} = -\ri (\omega - \omega_{a/b}) + \kappa_{a/b}$ and the diagonal matrix $\mathds{Q}_\pm = {\rm diag} \bigl( 1 , \re^{\pm \ri \theta}\bigr)$. This compact form makes obvious that we have an infinite set of equations in frequency space due to the coupling to the sidebands $\pm \Omega$, $\pm 2 \Omega$, etc.\ introduced by the modulation. These sidebands can be understood as generated by an electric synthetic field along the synthetic frequency axis (see Fig.~\ref{Fig:Sketch}). Furthermore, the phase shift itself can be interpreted by a synthetic magnetic field~\cite{PetersonEtAl2019} which adds a phase $\mathds{Q}_+$ for ``upward'' and $\mathds{Q}_-$ for ``downward'' transitions in the frequency bands. Recently, it has been shown theoretically and experimentally that this synthetic magnetic field results in non-reciprocal energy transmission in coupled oscillator systems~\cite{PetersonEtAl2019}. From the mathematical expression of $\mathds{Q}_\pm$ it is clear that $\mathds{Q}_+ = \mathds{Q}_-$ for phases $\phi = l \pi$ for all integers $l$. Hence for such phases the synthetic magnetic field makes no difference for ``upward'' and ``downward'' transitions and we can expect that there is no breaking of detailed balance by the synthetic magnetic field.

Furthermore, the compact matrix form allows us to write down formally the infinite set of equations in frequency space. To this end, we introduce the infinitely large block vectors
\begin{align}
	\uline{\boldsymbol{\psi}} &= (\ldots, \boldsymbol{\psi}_{++},\boldsymbol{\psi}_+, \boldsymbol{\psi}, \boldsymbol{\psi}_-, \boldsymbol{\psi}_{--}, \ldots)^t \\
	\uline{\mathbf{F}}  &=  (\ldots, \mathbf{F}_{++}, \mathbf{F}_+, \mathbf{F}, \mathbf{F}_-, \mathbf{F}_{--}, \ldots)^t 
\end{align}
where the indices are defined as $\mathbf{F}_\pm=\bigl(F_a(\omega \pm \Omega), F_b(\omega \pm \Omega) \bigr)^t$, $\mathbf{F}_{++/--} = \bigl(F_a(\omega \pm 2 \Omega), F_b(\omega \pm 2\Omega) \bigr)^t$, etc. Then we can rewrite the coupled Langevin Eqs.~(\ref{Eq:coupledLangevin}) as 
\begin{equation}
	\uline{\boldsymbol{\psi}} = \uline{\mathds{L}}^{-1} \uline{\mathds{M}} \uline{\mathbf{F}}
	\label{Eq:PerturbationSeries}
\end{equation}
where the diagonal and tridiagonal block matrices $\uline{\mathds{M}}$ and  $\uline{\mathds{L}}$ are defined in Ref.~\cite{Suppl}.
For any solution of this matrix equation it is necessary to consider only a finite subset. As typically done in such a Floquet-Shirley type approach we consider only block vectors of size $2(2N + 1)$ with the corresponding block matrices of size $2(2N + 1)\times2(2N + 1)$ centered around the solution for the zeroth sideband. The result can be considered as a perturbation result up to order $N$. %Hence, by considering subsequently larger matrices and vectors the solution will converge to the correct full solution. %For the numerical calculation we will later consider order $N = 20$ even though in most cases already lower orders will give a sufficiently convergent result.

Finally, we can derive a general expression for the spectral correlation functions $\langle a^\dagger a \rangle_\omega$, $\langle b^\dagger b \rangle_\omega$, $\langle a^\dagger b \rangle_\omega$, and $\langle b^\dagger a \rangle_\omega$. To this end, we first separate the contributions to $\uline{\boldsymbol{\psi}}$ due to the bath operator $\uline{\mathbf{F}}_a$ and $\uline{\mathbf{F}}_b$ by introduce the two block matrices $\uline{\mathds{Y}}_a = {\rm diag}(1,0,1,0, \ldots)$ and $\uline{\mathds{Y}}_b = {\rm diag}(0,1,0,1, \ldots)$ so that $\uline{\mathds{Y}}_a + \uline{\mathds{Y}}_b = \uline{\mathds{1}}$. These two matrices allow us to split the contributions from bath $a$ and bath $b$ so that
\begin{equation}
	\uline{\boldsymbol{\psi}} = \uline{\mathds{L}}^{-1} \uline{\mathds{M}} \mathds{Y}_a \uline{\mathbf{F}} + \uline{\mathds{L}}^{-1} \uline{\mathds{M}} \mathds{Y}_b \uline{\mathbf{F}}.
\end{equation}
We assume that the bath operators fulfill the fluctuation-dissipation theorem in the form ($i,j = a,b$)
\begin{equation}
	\langle F_{i}^\dagger (\omega + l \Omega) F_{j}(\omega' + l' \Omega) \rangle = \delta_{i,j} \delta_{l,l'} 2 \pi \delta(\omega - \omega') \langle F_i^\dagger F_i \rangle_\omega,
\end{equation}
with $\langle F_a^\dagger F_a \rangle_\omega  = 2 \kappa_a n_a(\omega_a)$ and $\langle F_b^\dagger F_b \rangle_\omega  = 2 \kappa_b n_b(\omega_b)$. 
%where the mean occupation numbers of the oscillators  
%\begin{equation}
%	n_{a/b} = \frac{1}{\re^{\hbar \omega_{a/b}/ \kb T_{a/b}} - 1}
%\end{equation}
%are determined by the oscillator frequencies $\omega_{a/b}$ and the bath temperatures $T_{a/b}$.  
This means that we assume that both baths are uncoupled and that the bath spectra are white noise spectra. This is a good assumptions as long as all important sidebands are close to $\omega_{a/b}$. This assumption can be made as long as the modulation amplitudes and the modulation frequency are small enough. Therewith we arrive at the final result 
\begin{equation}
%\begin{split}
	\langle	\uline{\boldsymbol{\psi}}_\alpha^\dagger \uline{\boldsymbol{\psi}}_\beta \rangle_\omega = \sum_{j = a,b} 2 \kappa_j n_j(\omega_j) \bigl( \uline{\mathds{L}}^{-1} \uline{\mathds{M}} \uline{\mathds{Y}}_j  \uline{\mathds{M}}^\dagger {\uline{\mathds{L}}^{-1}}^\dagger \bigr)_{\beta,\alpha} 
%	&\quad + 2 \kappa_b n_b(\omega_b) \bigl( \uline{\mathds{L}}^{-1} \uline{\mathds{M}} \uline{\mathds{Y}}_b \uline{\mathds{M}}^\dagger {\uline{\mathds{L}}^{-1}}^\dagger \bigr)_{\epsilon,\alpha}
%\end{split}
	\label{Eq:FullSpectrum}
\end{equation}
using the properties $\uline{\mathds{Y}}_{a/b}^\dagger = \uline{\mathds{Y}}_{a/b}$ and $\uline{\mathds{Y}}_{a/b} \uline{\mathds{Y}}_{a/b} = \uline{\mathds{Y}}_{a/b}$. From this expression we can numerically calculate all spectral correlation functions. For instance $\langle a^\dagger a \rangle_\omega$ is given by the component  $\alpha = 2N + 1$  and $\beta = 2 N + 1$, $\langle a^\dagger b \rangle_\omega$ by the component $\alpha = 2N + 1$ and $\beta = 2 N + 2$, etc. Note, that the such obtained spectral correlation functions are the sum of all sideband frequency components with equal weighting factors $2 \kappa_a n_a$ and $2 \kappa_b n_b$
.

\begin{figure}
	\centering
	  \includegraphics[width=0.45\textwidth]{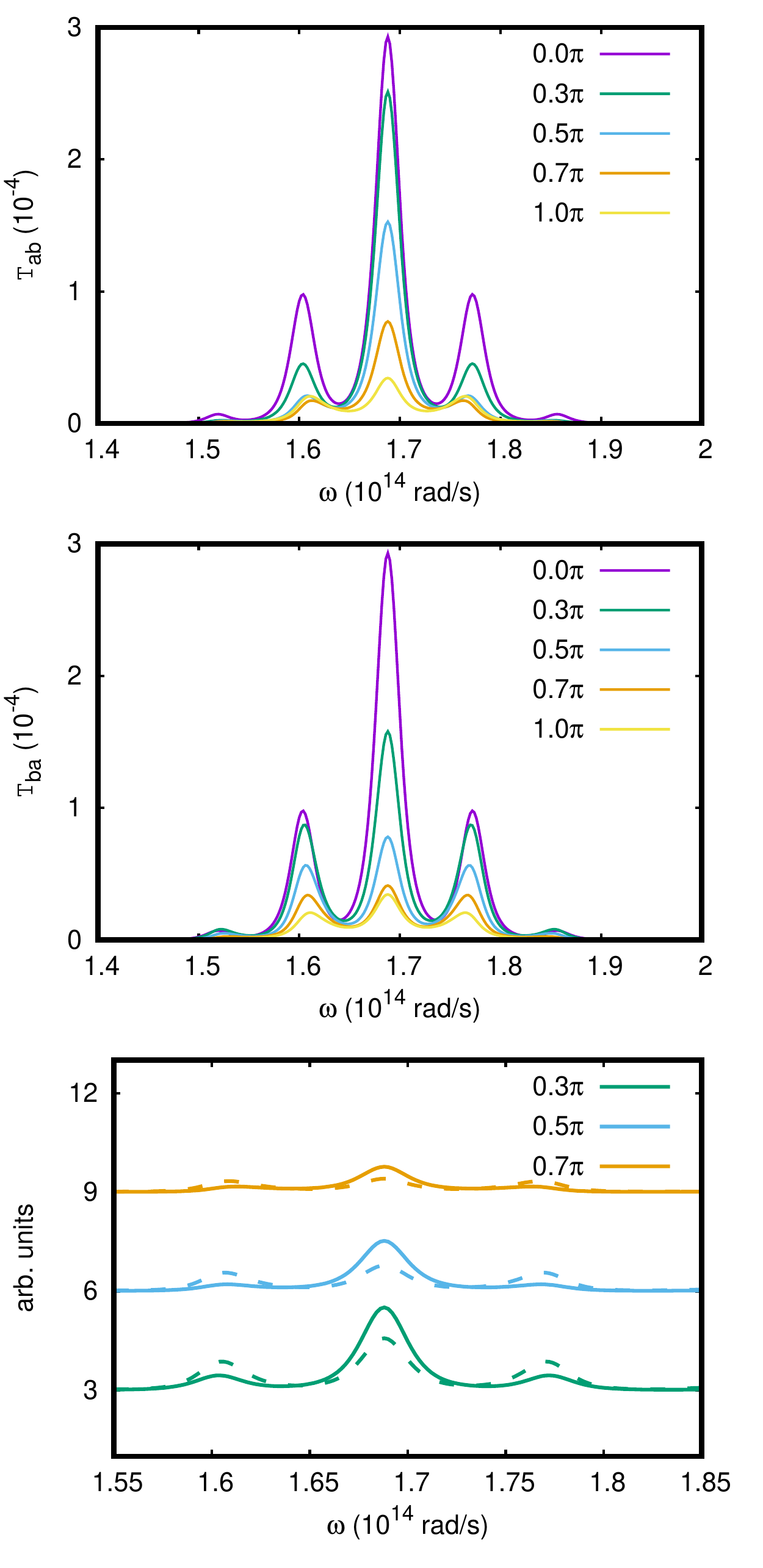}
	\caption{\label{Fig:flux} Non-reciprocal HTF $\mathcal{T}_{ab/ba}$ for graphene flakes at distance $d = 100\,{\rm nm}$ using perturbation order $N = 20$. Top: $\mathcal{T}_{ab}$ for $n_a \neq 0$ and $n_b = 0$. Middle: $\mathcal{T}_{ba}$ for $n_b \neq 0$ and $n_a = 0$. Parameters: $\beta = 0.05\omega_p$, $\Omega = 0.05\omega_p$, and $\theta$ is varied. Bottom: $\mathcal{T}_{ab}$ (full lines) and  $\mathcal{T}_{ba}$ (dashed lines) for $\theta = 0.3\pi, 0.5\pi, 0.7\pi$ with zero lines shifted to 3, 6, 9, resp.}
\end{figure}

Let us now use the model to discuss the heat flux between two graphene flakes. Graphene flakes have sharp resonances like plasmonic nanoparticles. The permittivity
of a graphene flake lying within a plane parallel to the x-y plane is given by the polarizability tensor $\uuline{\alpha} = {\rm diag}(\alpha, \alpha, 0)$ with~\cite{Graphene}
\begin{equation}
   \alpha  = \frac{3 c^3 k_r}{2 \omega_p^2} \frac{1}{\omega_p^2 - \omega^2 - \ri k \omega}
\end{equation}
with plasma frequency $\omega_p$, amplitude $k_r$, and damping constant $k$ which depent on the Fermi energies $E_F$ (in eV)~\cite{Suppl} and can be changed electrically or optically so that a modulation of the resonance frequency is feasible~\cite{Zwol,Minnich2}. The HTF between two identical graphene flakes facing each other at a distance $d$ is within fluctuational electrodynamics in the quasi-static regime given by~\cite{Suppl}
\begin{equation}
	\mathcal{T}_{ab}(\omega) = \mathcal{T}_{ba}(\omega) = 8 \frac{(\alpha'')^2}{(4 \pi d^3)^2}\frac{1}{|1 + \frac{\alpha^2}{(4 \pi d^3)^2}|^2}.
	\label{Eq:Tquasistatic}
\end{equation}
This HTF can now be related to our model. Within our model, by setting $F_b = 0$ the steady state power from $a$ to $b$ is~\cite{Suppl} 
\begin{equation}
	P_{a \rightarrow b} = \int_0^\infty \frac{\rd \omega}{2 \pi} \hbar \omega_a 2 \kappa_a \langle b^\dagger b \rangle_\omega
\end{equation}
so that the HTF is 
\begin{equation}
	\mathcal{T}_{ab} = \frac{2 \kappa_a}{n_a(\omega_a)} \langle b^\dagger b \rangle_\omega.
\end{equation}
Similarly, $\mathcal{T}_{ba}$ can be obtained by exchanging $a$ and $b$. Without any modulation we can directly determine the HTF from Eq.~(\ref{Eq:coupledLangevin}) for the case $\beta = 0$ and taking $F_b = 0$. Then we obtain~\cite{Suppl}
\begin{equation}
	\mathcal{T}_{ab} = \frac{4 (g \kappa_a)^2}{|X_a^2 + g^2|^2}. 
\label{Eq:SpektrumOscillator}
\end{equation}
By identifying the resonance frequency $\omega_a = \omega_b$ with the plasma frequency $\omega_p$ and $\kappa_a = \kappa_b \equiv \kappa$ with damping constant $k$ of the graphene sheet we can fit the HTF of our model to that of Eq.~(\ref{Eq:Tquasistatic}). We obtain a very good spectral fit for $g = 0.011\kappa_a$ (see Supplemental Material~\cite{Suppl}).

\begin{figure}
	\centering
	  \includegraphics[width=0.45\textwidth]{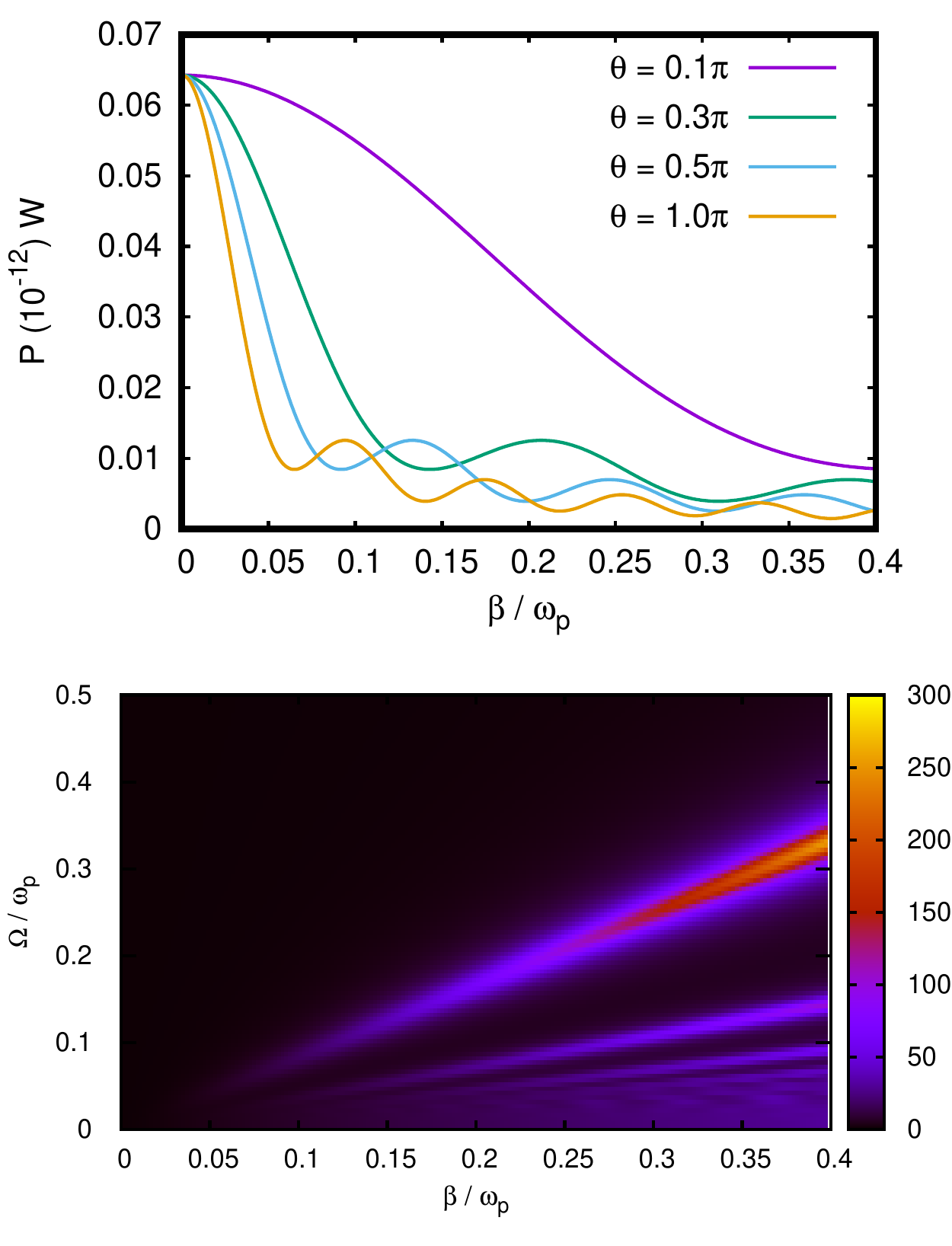}
	 \caption{\label{Fig:HF} Top: Total heat flux $P_{a \rightarrow b}$ between two graphene flakes at distance $d = 100\,{\rm nm}$ ($g/\kappa = 3.9$) for $T_a = 300\,{\rm K}$ and $T_b = 0\,{\rm K}$ as function of the modulation amplitude $\beta$ using dephasings $\theta = 0.1\pi, 0.3\pi, 0.5\pi, 1.0 \pi$. Bottom:  $P_0 / P_{a \rightarrow b}$ as function of $\beta$ and $\Omega$ where $P_0$ is the value without modulation for $\theta = 1.0\pi$. Numerical calculation is done for $N = 20$.}
\end{figure}

In Fig.~\ref{Fig:flux} we show the numerical results for the HTF for two identical graphene sheets with $\omega_a = \omega_b = \omega_p = 1.69\times10^{14}\,{\rm rad/s}$ and $\kappa_a = \kappa_b = 0.013\omega_p$ for $E_F = 0.4\,{\rm eV}$ when the resonance frequencies are modulated as in Eqs.~(\ref{Eq:omegaa}). We choose a relatively small amplitude $\beta = 0.05\omega_p$ which approximately corresponds to a change of the Fermi energy by $0.05\,{\rm eV}$ and a relatively small modulation frequency $\Omega = 0.05\omega_p$. First of all it can be seen that as expected the modulation produces side bands around the resonance frequency $\omega_p$. More interesting is that the spectra are in general different for $\theta \neq l \pi$ ($l \in \mathbf{Z}$) so that the detailed balance between the HTF is broken and we clearly have $\mathcal{T}_{ab}(\omega) \neq \mathcal{T}_{ba}(\omega)$ which is clearly due to the synthetic electric and magnetic fields. However, the integrated heat flux shown in Fig.~\ref{Fig:HF} is reciprocal so that we find $P_{a \rightarrow b} = P_{b \rightarrow a}$ which is equivalent to saying that the integrals over $\mathcal{T}_{ab}$ and $\mathcal{T}_{ba}$ are exactly the same which is in full agreement with the general statement in Eq.~(\ref{Eq:Integral}). Another feature is that the heat flux can be inhibited due to the modulation which can be easily understood by the fact that the resonances do less overlap during a modulation cycle when they are phase shifted. It is an interesting feature that this inhibition can be extremely high for specific combinations of $\Omega$ and $\beta$ in particular for $\beta \approx \Omega$ the heat flux can be up to 300 times smaller than without modulation for moderate values of modulation frequency and amplitude.  

Next, we consider a system of two membranes coupled by Casimir forces which allow for measurements of the spectra of the mean occupation numbers $\langle a^\dagger a\rangle_\omega$ and $\langle b^\dagger b \rangle_\omega$ of the membranes as done in Ref.~\cite{FongEtAl}, for instance. In that work the parameters are given by $\kappa_a = \omega_a/(2 Q_a)$ and  $\kappa_b = \omega_b/(2 Q_b)$ for the damping with oscillation frequencies $\omega_a = \omega_b = 2 \pi \cdot 191.6\,{\rm kHz} \equiv \omega_0$ and quality factors $Q_a = 4.5\times10^{4}$ and $Q_b = 2\cdot10^4$. For an unambiguous identification of the impact of the synthetic fields we choose $\kappa_a = \kappa_b = 10 \omega_a/(2 Q_a) \equiv \kappa$ which is much larger than in the actual experiment. The coupling constant due to the Casimir force is $g(d) = d^{-4.91} 2\times10^{-30} \, s^{-1}$. The measurements were carried out for distances from $d = 300\,{\rm nm}$ which is in the strong coupling regime ($g / \kappa = 1.54$) to $d = 800\,{\rm nm}$ in the weak coupling regime ($g / \kappa_1 = 0.013$). The transition between both regimes ($g / \kappa = 1$) occurs at a distance of about 330nm .

\begin{figure}
	\centering
	  \includegraphics[width=0.45\textwidth]{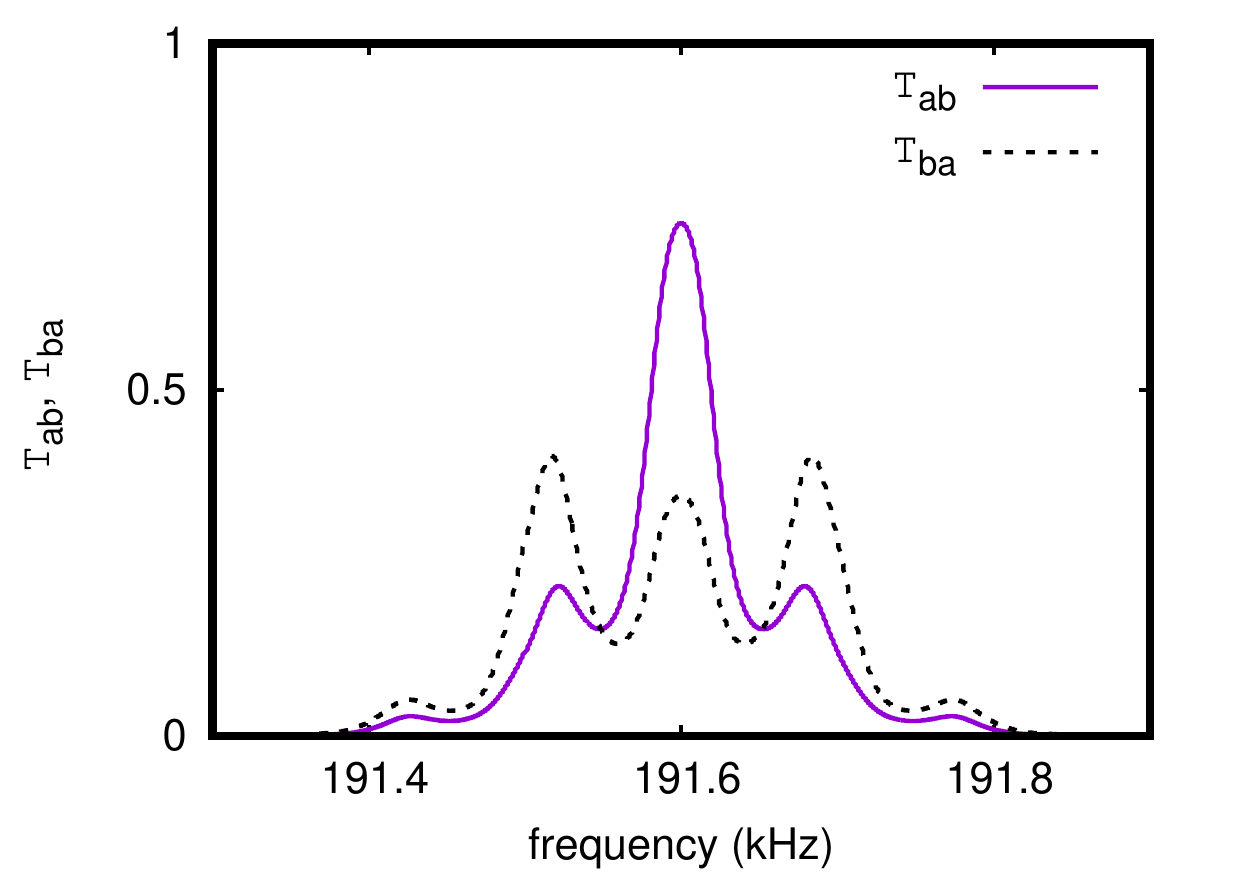}
	 \caption{\label{Fig:StrongCouplingImbalance} HTFs $\mathcal{T}_{ab}$ (full lines) and  $\mathcal{T}_{ba}$ (dashed lines) of two coupled membranes in strong coupling regime with $d = 300\,{\rm nm}$ with modulation parameters $\Omega = 0.0005\omega_0$,  $\beta = 0.0005\omega_0$, and $\theta = \pi/2$. Numerical calculation is done for $N = 10$.}
\end{figure}

In Fig.~\ref{Fig:StrongCouplingImbalance} we show the HTFs $\mathcal{T}_{ab}$ and $\mathcal{T}_{ba}$ in the strong-coupling regime. The broken detailed balance in the strong-coupling can be nicely seen. For an experimental verification a measurement of the mean occupation numbers as done in Ref.~\cite{FongEtAl} can be made which show imbalances directly connected with the broken detailed balance (see Fig.~3 and 4 in~\cite{Suppl}). However, when assuming that $n_a = n_b$ we find that  $\langle a^\dagger a \rangle / n_a + \langle a^\dagger a \rangle / n_b $ equals exactly $\langle b^\dagger b \rangle / n_a + \langle b^\dagger b \rangle / n_b$. Hence in global equilibrium the synthetic field has no impact on the total occupation numbers of the membranes which coincides with the result that the heat flux is reciprocal, i.e.\ $P_{a \rightarrow b} = P_{b \rightarrow a}$. Therefore the impact of the synthetic fields can only be measured when both membranes have different temperatures as realized in Ref.~\cite{FongEtAl}. It has to be emphasized that the broken detailed balance due to the synthetic fields as seen in the imbalance of the occupation numbers becomes prominent in strong-coupling regime~\cite{Suppl}.

In conclusion, we have shown that the presence of electric and magnetic synthetic fields breaks the detailed balance of the HTF but without resulting into a non-reciprocal total heat flux between two objects. We have discussed this phenomenon for the near-field radiative heat transfer between two graphene flakes. Furthermore we could show that synthetic fields allow for a strong heat flux inhibition which can be used to thermally isolate the graphene flakes by periodic modulations. Finally, we propose to measure the breakdown of detailed balance by measuring the mean occupation numbers of Casimir forced coupled membranes having different temperatures as recently done without dynamic modulation.

\acknowledgements

S.-A.\ B.\ acknowledges helpful discussions with Philippe Ben-Abdallah and support from Heisenberg Programme of the Deutsche Forschungsgemeinschaft (DFG, German Research Foundation) under the project No.\ 404073166. This research was supported in part by the National Science Foundation under Grant No. NSF PHY-1748958. G.S.\ A.\ thanks the kind support of The Air Force Office of Scientific Research [AFOSR award no. FA9550-20-1-0366], The Robert A. Welch Foundation [grant no. A-1943] and the Infosys Foundation Chair of the Department of Physics, IISc Bangalore.

\end{document}